\documentclass[10pt,journal,compsoc]{IEEEtran}
\ifCLASSOPTIONcompsoc
  \usepackage[nocompress]{cite}
\else
  \usepackage{cite}
\fi
\ifCLASSINFOpdf
   \usepackage[pdftex]{graphicx}
   \graphicspath{{../pdf/}{../jpeg/}}
  \DeclareGraphicsExtensions{.pdf,.jpeg,.png}
\else
   \usepackage[dvips]{graphicx}
   \graphicspath{{../eps/}}
\fi
\usepackage{amsmath}
\usepackage{color}
\usepackage[noend]{algpseudocode}
\usepackage{amssymb}
\usepackage{algorithmicx,algorithm}

\ifCLASSOPTIONcompsoc
  \usepackage[caption=false,font=footnotesize,labelfont=sf,textfont=sf]{subfig}
\else
  \usepackage[caption=false,font=footnotesize]{subfig}
\fi

\usepackage{stfloats}
\begin{document}
%
\title{Solving imperfect-information games via exponential counterfactual regret minimization}
%
%
%
%

\author{Huale Li, Xuan Wang, Shuhan Qi, Jiajia Zhang, Yang Liu, Yulin Wu, Fengwei Jia 
	\IEEEcompsocitemizethanks{\IEEEcompsocthanksitem 
		School of Computer Science and Technology,
		Harbin Institute of Technology, Shenzhen, China,518055
		\protect\\
		E-mail: lihuale@stu.hit.edu.cn
	}
}

%
%

\markboth{Journal of IEEE Intelligent Systems}%
{Shell \MakeLowercase{\textit{et al.}}: Bare Demo of IEEEtran.cls for Computer Society Journals}
%



\IEEEtitleabstractindextext{%
\begin{abstract}
In general, two-agent decision-making problems can be modeled as a two-player game, and a typical solution is to find a Nash equilibrium in such game. Counterfactual regret minimization (CFR) is a well-known method to find a Nash equilibrium strategy in a two-player zero-sum game with imperfect information. The CFR method adopts a regret matching algorithm iteratively to reduce regret values progressively, enabling the average strategy to approach a Nash equilibrium. Although CFR-based methods have achieved significant success in the field of imperfect information games, there is still scope for improvement in the efficiency of convergence. To address this challenge, we propose a novel CFR-based method named exponential counterfactual regret minimization (ECFR). With ECFR, an exponential weighting technique is used to reweight the instantaneous regret value during the process of iteration. A theoretical proof  is provided to guarantees convergence of the ECFR algorithm. The result of an extensive set of experimental tests demostrate that the ECFR algorithm converges faster than the current state-of-the-art CFR-based methods.
\end{abstract}

}

\maketitle

\IEEEdisplaynontitleabstractindextext

%
\IEEEpeerreviewmaketitle

\IEEEraisesectionheading{\section{Introduction}\label{sec:introduction}}

%
%
%
%
\IEEEPARstart{G}{ame} theory has often been regarded as a touchstone to verify the theory of artificial intelligence \cite{Fudenberg1998The}. In general, games can be divided into perfect information games (PIGs) and imperfect information games (IIGs), according to whether the player has complete knowledge of the game state of all the other players. Examples of PIGs are Go and Chess, whereas games like poker are IIGs 
that players cannot observe the complete game state of the other players. In other words, compared with the PIGs, the IIGs usually holds certain private information. Recently IIGs have attracted a great deal of attention from researchers, since IIGs are much more challenging compared with PIGs. In this paper, we mainly focus on the study of decision-making for IIGs \cite{1997Game}.

In order to solve a two-player zero-sum game with imperfect information, a typical solution is to find a Nash equilibrium strategy \cite{nash1951non}. As a popular method of computing the Nash equilibrium strategy, counterfactual regret minimization (CFR) \cite{zinkevich2008regret} has attracted the attention of many researchers due to its sound theoretical guarantee of convergence. Over the past decade, many variants of CFR have been developed \cite{zinkevich2008regret,lanctot2009monte,bowling2015heads,johanson2012efficient,brown2019solving}. For example, Monte Carlo counterfactual regret minimization (MCCFR) is a sample-based CFR algorithm, which combines Monte Carlo methods with standard or vanilla CFR to compute approximate equilibria strategies in IIGs\cite{lanctot2009monte}. CFR+ uses a variant of regret matching (regret matching+) where regrets are constrained to be non-negative \cite{bowling2015heads}. Double neural CFR \cite{li2018double} and Deep CFR \cite{brown2019deep} combine deep neural networks with vanilla CFR and linear CFR (LCFR) respectively. Moreover, there are many other CFR variants, such as public chance sampling CFR (PCCFR) \cite{johanson2012efficient}, variance reduction in MCCFR (VR-MCCFR) \cite{schmid2019variance} and discounted CFR (DCFR) \cite{brown2019solving}. 

Moreover, the CFR-based methods have achieved notable success in the domain of IIGs, especially in poker games \cite{bowling2015heads,Brown2017Superhuman,moravvcik2017deepstack,brown2019superhuman}. Libratus is the first agent to beat top human players in heads-up no-limit Texas Hold'em poker \cite{Brown2017Superhuman}. DeepStack also defeated professional poker players, and its method has been demonstrated to be sound with a theoretical proof \cite{moravvcik2017deepstack}. Pluribus, the latest computer program based on CFR, defeated top poker players in six-player no-limit Texas Hold'em poker, which can be recognized as a milestone in the field of artificial intelligence and game theory \cite{brown2019superhuman}. 

Although the agents based on the CFR method have achieved conspicuous success, they are not directly improving the vanilla CFR method itself. In other words, these agents are operated by combining CFR with other techniques (abstraction, neural network, \textit{etc}.) to solve the game strategy. Moreover, vanilla CFR is an iterative strategy solving method that relies on increasing the number of training iterations to improve the accuracy of the strategy. However, this means that finding a robust strategy may require a large number of iterations, which leads a significant amount of time for solving. This makes CFR-based methods difficult to be applied in some real situations. Therefore, it is worth investigating how to accelerate the convergence of the CFR and obtain a robust strategy more efficiently. 

Towards this goal, we propose an exponential CFR (ECFR), which speedups the CFR by focusing more attention on the actions with greater advantage. Actually, the regret value of an action represents the advantage of this action compared with the expected action of current strategy. Therefore, we utilize average regret value as a threshold to filter the advantage action. Then, the action with different advantage will be weighted differently, \textit{i.e.}, the actions with higher advantage will be given higher weight, and vice versa. In this way, the updated strategy will have a tendency to seclect the advantage action. And finally, the convergence of CFR is accelerated.

To be specific, we adopt regret value as the metric to evaluate the advantage of actions. We propose an exponential weighting technique that applies an exponential weight to actions with a higher regret. This causes the iterative algorithm to pay more attention on actions with higher regret values and so accelerates  the training process. Also, in contrast to traditional methods, actions with negative regret values are also considered, instead of setting their probabilities to zero. We summarize our contributions as follows:
\begin{itemize}
	\item [1)]  We present an exponential weighting technique that is applied to vanilla CFR, called exponential CFR (ECFR), which makes the convergence of CFR more efficiently. 
	\item [2)] We give a proof of convergence for our ECFR, which, like CFR, provides it with a solid theoretical basis. 
	\item [3)] Three different games (Kuhn poker, Leduc poker and Royal poker) are used to evaluate our ECFR. Extensive experimental results show that ECFR converges faster in the three kinds of games compared with the current state-of-the-art CFR-based methods.
\end{itemize}

The rest of our paper is organized as follows. In Sect.2 we introduce the concepts of extensive form game, describe Nash equilibrium, and CFR. Our proposed method is described in Sect.3, which includes details of the ECFR and its proof of convergence. In Sect.4, we evaluate the performance of the ECFR on four varieties of IIG. Finally, in Sect.5, we present the conclusions of the paper.
\section{Notation and Preliminaries}
In this section, some notation and definitions of extensive-form games are introduced. Then, the concept of Nash equilibrium is described. Finally, an overview of CFR is provided.
\subsection{Extensive-Form Game Model}
\begin{figure}[!t]
	\centering
	\includegraphics[width=3.5in]{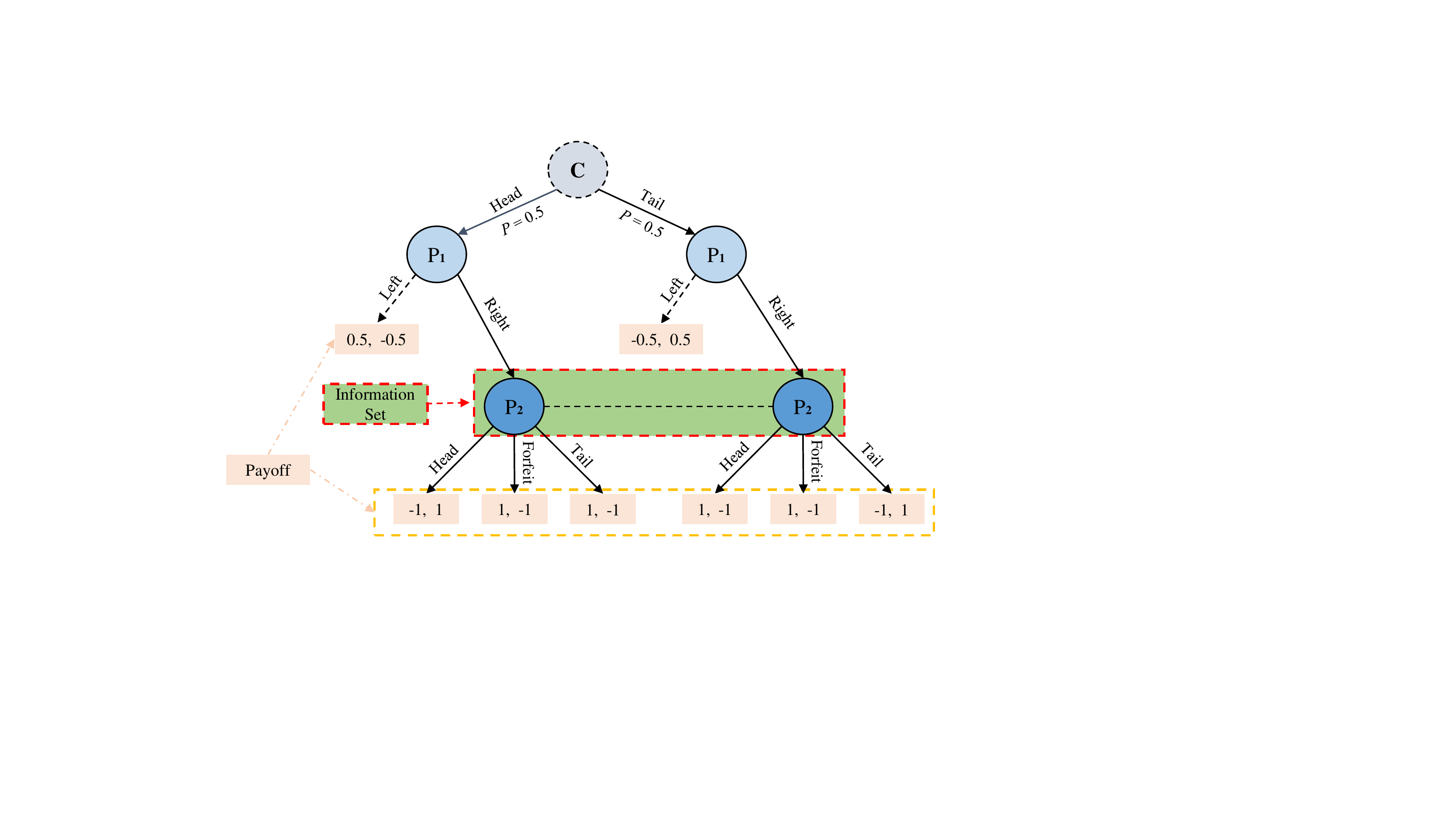}
	\caption{The game tree of the  Coin Toss game. ("C" represents a chance node. $P_1$ and $P_2$ are game players. A coin is flipped and lands either Heads or Tails with equal probability, but only player $P_1$ can see the outcome. The information set of $P_2$ is the dotted line between the two $P_2$ nodes, which means that player $P_2$ cannot distinguish between the two states.)}
	\label{fig1}       
\end{figure}

The extensive-form game is a classical model for sequential decision-making. It is particularly useful in the field of IIGs, in which some decision-makers have private information to each other. Before presenting the formal definition, we consider a specific example of the Coin Toss game. Fig.~\ref{fig1} represents a game tree for the game of Coin Toss.

As shown in Fig.~\ref{fig1}, each node represents a game state within the game tree. A leaf node, which is also known as a terminal node, indicates that the game has ended. The corresponding payoff is returned after the game ends. In addition, the edge between two nodes represents an action or decision taken by a game player. A coin is flipped and lands either Heads or Tails with equal probability in Coin Toss game, but only player $P_1$ knows the outcome. In Fig.~\ref{fig1}, player $P_1$ can choose between the actions Left and Right, with the action Left leading directly to obtaining a payoff. If the action Right is selected by player $P_1$, then player $P_2$ has an opportunity to guess how the coin landed. If $P_2$ guesses correctly, $P_1$ will receive a reward of -1 and $P_2$ will receive a reward of 1 \cite{Brown2017Safe}.

Generally, a finite extensive-form game with imperfect information has six components, represented as $\left<N,H,P,f_{c},I,u\right>$ \cite{Martin1994}: $N$ represents the game players. $H$ is a limited set of sequences that represent the possible historical actions. \textit{P} is the player function. $P(h)$ is the player who takes action $a$ after history $h$. If $P(h)=c$, then chance determines the action after history $h$. $f_c$ is a function that associates every history $h$, and $f_{c}(a\mid h)$ is the probability that action $a$ occurs given history $h$. $I$ is the information set. For any information set $I_i$ belonging to player $i$, all nodes $h$, $h^{\prime} \in I_{i}$ are indistinguishable to player $i$. \textbf{$u$} is a utility function for every termination state. Due to the number of symbols in this paper, we give a brief list of variables for further reference in Table ~\ref{tab:1}.
\begin{table}[!t]
	\renewcommand{\arraystretch}{1.3}
	\caption{Description of variables}
	\label{tab:1}       
	\centering
	\begin{tabular}{|c||l|}
		\hline  
		\textbf{Variable}&	\textbf{Meaning} \\
	\hline \hline
		\textbf{$u$}& the utility function where $u_i$ represents the utility of  \\&
		 player $i$\\
		 \hline
		$H$ &  a limited set of sequences, which is the possible set of \\& historical actions \\
		\hline
		$\sigma$&the strategy where $\sigma_i$ is the strategy of player $i$, and $\sigma_{-i}$ is \\&the strategy of the other player\\
		\hline
		$\mathcal{I}_{i}$&	the information set of player  $i$\\
		\hline
		$\pi^{\sigma}(h)$& the joint probability of reaching $h$ if all players play\\& according to $\sigma$, $\pi_i^\sigma(h)$ is the probability of reaching  $h$ \\&if player $i$ operates according to $\sigma$\\
		\hline
		
	\end{tabular}
\end{table}

\subsection{Nash Equilibrium}
The Nash equilibrium is a fundamental concept in game theory, which lays the theoretical foundation for many studies. A Nash equilibrium is usually used to computed the strategy of a two-player extensive-form game in the field of the IIG, which can be also called a non-cooperative game equilibrium (NE) \cite{nash1951non}. To better understand the Nash equilibrium, here we introduce the concept of a strategy as follows.

Strategy $\sigma$ is a probability vector over actions in the extensive-form game, where $\sigma_i$ is the strategy of player $i$. $\sigma_i(I,a)$ represents the probability of player $i$ taking action $a$ under the information set $I$. $\sigma_{-i}$ refers to all the strategies in $\sigma$ except player $i’$'s strategy $\sigma_i$. $\pi^{\sigma}(h)$ is the probability of history $h$ that occurs only if the game player takes the legal actions according to strategy $\sigma$. $\pi_{-i}^{\sigma}(h)$ of a history $h$ is the contribution of chance and all players other than the player $i$. $u_{i}\left(\sigma_{i}, \sigma_{-i}\right)$ is the expected payoff for player $i$ if all players play according to the strategy profile ($\sigma_i,\sigma_{-i}$).

A best response to $\sigma_{-i}$ is a strategy of player $i$, $BR(\sigma_{-i})$ such that $u_{i}\left(B R\left(\sigma_{-i}\right), \sigma_{-i}\right)=\max _{\sigma_{i}^{\prime}} u_{i}\left(\sigma_{i}^{\prime}, \sigma_{-i}\right)$. A Nash equilibrium $\sigma^*$ is a strategy profile where everyone plays a best response: $\forall i, u_{i}\left(\sigma_{i}^{*}, \sigma_{-i}^{*}\right)=\max _{\sigma_{i}^{\prime}} u_{i}\left(\sigma_{i}^{\prime}, \sigma_{-i}^{*}\right)$ \cite{nash1951non}. 

\subsection{Counterfactual Regret Minimization}\label{seccfr}
Counterfactual regret minimization (CFR) is a popular iterative method to find the Nash equilibrium strategy in two-player zero-sum games with imperfect information \cite{zinkevich2008regret}. In general, CFR can be divided into two steps: regret calculation and regret matching. The former one is to calculate the regret values of actions in each iteration. And the latter one is to update global strategy. We provide an overview of vanilla CFR as the following. 

\textbf{Step 1}: Let $\sigma_t$ be the strategy at iteration $t$. The instant regret $r^t(I,a)$ on iteration $t$ for the action $a$ in the information set $I$ is formally defined as follows:
\begin{equation}\label{eq1}
 r^{t}(I, a)=v^{\sigma^{t}}(I, a)-v^{\sigma^{t}}(I)
\end{equation}
where $v^{\sigma}(I)$ is conterfactual value, which represents the expected payoff of player $i$. $v^{\sigma}(I)$ is weighted by the probability that the palyer $i$ would reached $I$ if the player tried to do so that iteration when reaching the information set $I$. The conterfactual value $v^{\sigma}(I)$ is defined as:
\begin{equation}
v^{\sigma}(I)=\sum_{h \in I}\left(\pi_{-i}^{\sigma}(h) \sum_{z \in \mathcal{Z}}\left(\pi^{\sigma}(h, z) u_{i}(z)\right)\right)
\end{equation}
where $u_i(z)$ is the payoff of the player $i$ in the leaf node $z$. And the counterfactual value of the action $a$ is defined as:
\begin{equation}
v^{\sigma}(I, a)=\sum_{h \in I}\left(\pi_{-i}^{\sigma}(h) \sum_{z \in \mathcal{Z}}\left(\pi^{\sigma}(h \cdot a, z) u_{i}(z)\right)\right)
\end{equation}

The counterfactual regret $R^{T}(I, a)$ for the action $a$ in the information set $I$ on $T$ iterations is defined as:
\begin{equation}
R^{T}(I, a)=\sum_{t=1}^{T} r^{t}(I, a)
\end{equation}

\textbf{Step 2}: We define $R_{+}^{T}(I, a)=\max \left\{R^{T}(I, a), 0\right\}$. The CFR algorithm updates its strategy iteratively through the regret matching algorithm (RM) on each information set. In RM, a player picks a distribution over actions in an information set in proportion to the positive regret of those actions. Formally, on iteration $T+1$, the player selects actions $a \in A(I)$ according to the probabilities:
\begin{equation}
	\sigma^{T+1}(I, a)=\left\{\begin{array}{ll}
	\frac{R_{+}^{T}(I, a)}{\sum_{a^{\prime} \in A(I)} R_{+}^{T}\left(I, a^{\prime}\right)}, & \text { if } \sum_{a^{\prime}} R_{+}^{T}\left(I, a^{\prime}\right)>0 \\
	\frac{1}{|A(I)|}, & \text { otherwise }
	\end{array}\right.
\end{equation}

The regret matching algorithm is able to guarantee that its level of regret will decrease over time, so that it will eventually achieve the same effect as a Nash equilibrium strategy.
\section{Our Method}
In this section, firstly, the exponential weighting technique is presented in Sect. 3.1. Secondly, the process of ECFR is introduced in Sect. 3.2. Thirdly, the proof of convergence of ECFR is given in Sect. 3.3. Finally, the differences between our method and other CFR-based methods will be discussed.
\subsection{Exponential Weighting Technique}\label{ewt}
In recent years, there have been many improved methods based on vanilla CFR. For instance, Discounted CFR (DCFR) \cite{brown2019solving}, Linear CFR (LCFR) \cite{brown2019deep} and dynamic thresholding for CFR \cite{Noam2017dynamic} aim at speeding up the convergence of vanilla CFR.

Among these methods, both of LCFR and DCFR primarily balance the weight of regret generated in the early and later iterations. For LCFR \cite{brown2019deep} and DCFR \cite{brown2019solving}, their improvements over vanilla CFR are in their way of weighting the regret value. LCFR uses the number of iterations to modify the weight of the regret value. DCFR also reweights regret values, but this weight is different for positive and negative regret, which are $(t/(t+1))^\alpha$ and $(t/(t+1))^\beta$ respectively (where $t$ is the number of the iteration). \cite{Noam2017dynamic} introduced a dynamic thresholding for CFR in which a threshold is set at each iteration such that the probability of actions below the threshold is set to zero. 

Actually, the essence of CFR is that it is an iterative strategy, so it becomes progressively more accurate with an increase in iterations. The strategy affects the regret value, which decreases as the number of iterations increases. Therefore, the ultimate goal of both the vanilla CFR and several of its improved variations is to accelerate convergence, that is, to improve the speed of convergence. In this paper, following with the same purpose, we propose an exponential weighting technique, which tries to make the strategy converge faster by reweighting the regret value.

In CFR-based methods, different values for regret indicate that the corresponding level of importance is different. The regret value of an action represents the advantage of this action compared with the expected action of current strategy. This means it would be beneficial to focus on actions with higher regret values by giving them higher weights. 

In the method, an exponential weighting technique is proposed, which applies an exponential weight to actions with a higher level of regret. This makes the iterative algorithm pay more attention to actions that incur a higher regret and so improves the strategy accordingly.
 Its formal description is as follows:
\begin{equation}\label{exweight}
	f(x) = \left\{\begin{array}{ll}
	e^\alpha x, & \text { if } x>0 \\
	e^\alpha\beta, & \text{ if }  x\leq 0
	\end{array}\right.
\end{equation}
where $\alpha$ is a parameter that is closely related to the variable $x$, $\beta$ is a parameter with a small value that will be discussed in the Experiment section below, and $f(x)$ is the output. 

To be more specific, the variable $x$ of Eq.6 may be a negative value in the process of solving games with CFR. In contrast to conentional methods that set the negative value to 0, in our method we set the variable with a negative value to a new minimum value $e^\alpha*\beta$. This is because the strategy obtained in the early stages is not yet accurate enough, and some actions with a negative regret value are still worth considering. In the early stages of the strategy iteration, it is unreasonable to ignore actions with a negative regret when updating strategies. Therefore, the variable with a negative value is set to a new minimum value in our method.

\subsection{Exponential Counterfactual Regret Minimization}
We propose a novel CFR-based variant, known as Exponential Counterfactual Regret Minimization (ECFR). Our method is based on vanilla CFR, which redistributes the weight of instantaneous regret values through the exponential weighting technique introduced in the last section. 

As indicated in Eq.\ref{exweight}, the parameter $\alpha$ in the exponential weighting method is closely related to the variable $x$. Specifically in ECFR, we define a loss function, which can be regarded as the parameter $\alpha$ in the Eq.6. In addition, we regard the instantaneous regret value on each iteration as the variable $x$ in the exponential weighting technique. The instant regret $r_i^t(I,a)$ is filtered with the mean value $EV_I$, which makes the strategy of next iteration focuses on the more advantageous actions by giving them higher weights. The loss function can then be defined as follows:
\begin{equation}
L_1 = r_i^t(I,a) - EV_I
\end{equation}
where $ r_i^t(I,a) $ has the same definition as Eq.\ref{eq1} in Sect.\ref{seccfr} that is, the immediate regret value of action $a$ for player $i$ on iteration $t$. $EV_I$ is the average counterfactual regret value on each iteration, $EV_I = \frac{1}{|A(I)|} \sum_{a \in A(I)} r(I,a)$, $A(I)$ represents the legal actions on information set $I$. 

Following the approach of vanilla CFR, at each iteration, ECFR aims to minimize the total regret value by minimizing the regret on each information set. In contrast to vanilla CFR, ECFR uses a particular weight for the calculation of the immediate regret value. With iterations increase, ECFR pays more attention on the actions with higher instant regret values by introducing $L_1$ loss, which is weighted in exponential form. The regret for all actions $a\in A(I) $ on each information set $I$, $R(I,a)$ can be calculated as follows:
\begin{equation}\label{regret}
R_{i,ECFR}^T(I, a)=\left\{\begin{array}{ll}
{\sum_{t=1}^{T} e^{L_1}r_{i}^{t}(I, a),} & { \text{ if } \  r_i^t(I, a)>0} \\ {\sum_{t=1}^{T} e^{L_1}\beta,} & \text{ if }\ r_i^t(I, a) \leq 0
\end{array}\right.
\end{equation}
where $\beta$ is a parameter that will be set in the following section Sect.\ref{Ablation Study}.

Then the strategy for iteration $T+1$ can be computed with a regret matching algorithm (RM) as follows:
\begin{equation}\label{t+1}
\sigma_{i}^{T+1}(I,a)=\frac{e^{L_{1}} R_{i,ECFR}^{T}(I, a)}{\sum_{a^{\prime} \in A(I)} e^{L_{1}} R_{i,ECFR}^{T}\left(I, a^{\prime}\right)}
\end{equation}

In vanilla CFR, the total regret $R_{i}^{T} \leq \sum_{I \in \mathcal{I}_{i}} R^{T}(I)$ if player $i$ plays according to CFR on each iteration. Thus, as $T \rightarrow \infty$, then  $\frac{R_{i}^{T}}{T} \rightarrow 0$ \cite{zinkevich2008regret}. If the average regret of both players satisfies $\frac{R_{i}^{T}}{T} \leq \epsilon$, then their average strategy $\left\langle\bar{\sigma}_{1}^{T}, \bar{\sigma}_{2}^{T}\right\rangle$ will be a $2\epsilon$-Nash equilibrium in a two-player zero-sum game \cite{Waugh2009}, where the average strategy is updated as $\bar{\sigma}_{i}^{T}(I)=\frac{\sum_{t=1}^{T}\left(\pi_{i}^{\sigma^{t}}(I) \sigma_{i}^{t}(I)\right)}{\sum_{t=1}^{T} \pi_{i}^{\sigma^{t}}(I)}$.

For regret matching \cite{Sergiu2000A}, it proved that if $\sum_{t=1}^{\infty} w_{t}=\infty$ then the weighted average regret, which is defined as $R_{i}^{w, T}=\max _{a \in A} \frac{\sum_{t=1}^{T}\left(w_{t} r^{t}(a)\right)}{\sum_{t=1}^{T} w^{t}}$ is bounded by:
\begin{equation}\label{eq6}
	R_{i}^{w, T} \leq \frac{\Delta \sqrt{|A|} \sqrt{\sum_{t=1}^{T} w_{t}^{2}}}{\sum_{t=1}^{T} w_{t}}
\end{equation}

The work of \cite{Noam2014regret} has shown that the weighted average strategy $\sigma_{i}^{w, T}(I)=\frac{\sum_{t \in T}\left(w_{t} \pi_{i}^{\sigma^{t}}(I) \sigma_{i}^{t}(I)\right)}{\sum_{t \in T}\left(w_{t} \pi_{i}^{\sigma^{t}}(I)\right)}$ is a 2$\epsilon$-Nash equilibrium if the weighted average regret is $\epsilon$ in two-player zero-sum games. Our method can also obtain a similar theoretical guarantee that the average strategy of players computed with ECFR will eventually converge to a Nash equilibrium. The average strategy of the ECFR is as follows:
\begin{equation}\label{average}
\bar{\sigma}_{i}^{T}(I, a)=\frac{\sum_{t=1}^{T}e^{L_{1}} \pi_{i}^{\sigma^{t}}(I) \sigma^{t}(I, a)}{\sum_{t=1}^{T} e^{L_{1}}\pi_{i}^{\sigma^ t}(I)}
\end{equation}

The detailed proof of the convergence will be given in the next section. The ECFR is described in algorithm \ref{algorithm1} and algorithm \ref{algorithm2}. 

\begin{algorithm}[t]
	\caption{The ECFR} 
	\label{algorithm1}
	\hspace*{0.02in} {\bf Input:} 
	The game $G$, the strategy $\sigma^t_i$ for each player, the regret $r_i^t (I,a)$ and $R_i^t (I,a)$, $\beta$, iteration $T$.  \\
	\hspace*{0.02in} {\bf Output:} 
	The strategy $\sigma^{t+1}_i(I)$ of the next iteration, the average strategy $\bar{\sigma}_{i}^{T}(I)$. \\
    Initilize each player's strategy $\sigma^1_i$, the regret $r_i^1 (I,a)$ and $R_i^1 (I,a)$.
	\begin{algorithmic}[1]
		\For{ECFR iteration $t=1$ to $T$}
		\State Env = $G$
		\For{each player $i$}
		\State Traverse($\emptyset,i,\sigma^t_1(I),\sigma^t_2(I),t$)			
	    \EndFor
	    \State Compute the strategy of $t+1$ iteration $\sigma^{t+1}(I)$ with regret matching Eq.\ref{t+1}.
     	\EndFor
     	\State Compute the average strategy $\bar{\sigma}_{i}^{T}(I)$ with Eq.\ref{average}.
		\State \Return  $\sigma^{t+1}_i(I)$, $\bar{\sigma}_{i}^{T}(I)$
	\end{algorithmic}
\end{algorithm}

\begin{algorithm}[t]
	\caption{Traverse the game tree with ECFR} 
	\label{algorithm2}
	\textbf{function} Traverse($h,i,\sigma^t_1(I),\sigma^t_2(I),t$)  \\
	\hspace*{0.02in} {\bf Input:} 
	The history $h$, player $i$, strategies $\sigma^t_1(I)$ and $\sigma^t_2(I)$, ECFR iteration $t$.  
	\begin{algorithmic}[1]
	\If{$h$ is a terminal node} 
	\State \textbf{return} the payoff of the player $i$
	\ElsIf{$h$ is a chance node}
	\State $a \sim \sigma(h)$ 
	\State	\textbf{return} Traverse($h,i,\sigma^t_1(I),\sigma^t_2(I),t$)
	\ElsIf{$i$ is the acting player}  

	\For{$a\in A(h)$} 
	\State $v(I,a)\leftarrow$ Traverse($h,i,\sigma^t_1(I),\sigma^t_2(I),t$) \quad   $\vartriangleright$ Traverse each action
	\EndFor
	\For{$a\in A(h)$} 
	\State  $r^t(I,a) \leftarrow v(I,a)-\sum_{a'\in A(h)} \sigma^t(I,a')\cdot v(a')$  
	     
	\EndFor   
	\State Compute the regret $R_{i,ECFR}^T(I, a)$ with Eq.\ref{regret}
    \Else     \qquad \qquad \qquad    $\vartriangleright$ If the opponent is the acting player
    \State Compute the regret $R_{-i,ECFR}^T(I, a)$ with Eq.\ref{regret}
	\State \Return  Traverse($h,i,\sigma^t_1(I),\sigma^t_2(I),t$)
	\EndIf
	\end{algorithmic}
\end{algorithm}

\subsection{Proof of Convergence}
A brief but sufficient theoretical proof of convergence for ECFR is given in this section. As described in the previous section, our ECFR method proposes an RM based algorithm to adjust the average strategy in which the regret value is given different weights. Here we will prove that there is a convergence bound for the ECFR average strategy, and the bound is never higher than that of vanilla CFR.

\textbf{Theorem 1}  Assume that the number of iterations is $T$ and that ECFR is conducted as a two-player zero-sum game. Then the weighted average strategy profile is a $\frac{\Delta |\mathcal{I}| \sqrt{|A|}e^{2T-T^2}} {\sqrt{T}}$ Nash equilibrium.

\textit{Proof}. The lowest amount of instantaneous regret on any iteration is $-\Delta$. Consider the weighted sequence of iterations $\sigma^{\prime 1}, \ldots, \sigma^{\prime T}$, where $\sigma^{\prime t}$ is identical to $\sigma^t$, but the weight is $w_{a,t}=\prod_{i=t}^{T-1} e^i =e^\frac{(T+t-1)(T-t)}{2}$ rather than $w_{a,t}=\prod_{i=t}^{T-1} e^{L_1}$. $R^{\prime t}(I, a)$ is the regret of action $a$ on information set $I$ at iteration $t$, for this new sequence. 

In addition, for the regret matching, \cite{Cesa2006Prediction} proves that if $\sum_{t=1}^{\infty} w_{t}=\infty$, then the weighted average regret, defined as $R_{i}^{w, T}=\max _{a \in A} \frac{\sum_{t=1}^{T}\left(w_{t} r^{t}(a)\right)}{\sum_{t=1}^{T} w^{t} }$ is bounded by Eq.\ref{eq6} depicted in section 3.2.

We can find that $R^t(I,a) \leq \frac{\Delta\sqrt{|A|}e^{2T-T^2}}{ \sqrt{T}}$ for player $i$' performing action $a$ on information set $I$, from Lemma 3. We can use Lemma 1, which uses the weight $w_{\alpha, t}$ for iteration $t$ with $B= \frac{\Delta\sqrt{|A|}e^{2T-T^2}}{ \sqrt{T}}$ and $C=0$. This means that $R^{\prime t}(I, a) \leq w_T(B-C) \leq \frac{\Delta\sqrt{|A|}e^{2T-T^2}}{ \sqrt{T}} $ from Lemma 1.  Furthermore, we find that the weighted average regret is at most $\frac{\Delta\left|\mathcal{I}_{i}\right| \sqrt{|A|}e^{2T-T^2}}{ \sqrt{T}}$ from Lemma 3 because, for the information set,  $\left|\mathcal{I}_{1}\right|+\left|\mathcal{I}_{2}\right|=|\mathcal{I}|$, the weighted average strategies form a $\frac{\Delta |\mathcal{I}| \sqrt{|A|}e^{2T-T^2}} {\sqrt{T}}$-Nash
equilibrium (with the iteration $T$ increasing, $\frac{\Delta |\mathcal{I}| \sqrt{|A|}e^{2T-T^2}} {\sqrt{T}}$ approaches zero).

\textbf{Lemma 1}. Call a sequence $x_{1}, \ldots, x_{T}$ of bounded real values $B C$-plausible if $B>0, C \leq 0, \sum_{t=1}^{i} x_{t} \geq C$ for
all $i,$ and $\sum_{t=1}^{T} x_{t} \leq B .$ For any $B C$-plausible sequence and any sequence of non-decreasing weights $w_{t} \geq 0$,  $\sum_{t=1}^{T}\left(w_{t} x_{t}\right) \leq w_{T}(B-C)$.

\textbf{Lemma 2}. Given a group of actions $A$ and any sequence of rewards $v^{t}$, such that $\left|v^{t}(a)-v^{t}(b)\right| \leq \Delta$ for all $t$ and all $a, b \in A,$ after conducting a set of strategies decided by regret matching, apply the regret-like value $Q^{t}(a)$ instead of $R^{t}(a), Q^{T}(a) \leq \Delta \sqrt{|A| T}$ for all $a \in A$.

\textit{Proof}. This lemma closely resembles Lemma 1, and  both are from \cite{tammelin2015solving}, thus here we do not give the detailed proof of these two lemmas. 

\textbf{Lemma 3}. Suppose player $i$ conducts $T$ iterations based on ECFR, then the weighted regret for player $i$ is at most $\Delta\left|\mathcal{I}_{i}\right| \sqrt{|A|} \sqrt{T}$, and the weighted average regret for player $i$ is at most $\frac{\Delta\left|\mathcal{I}_{i}\right| \sqrt{|A|}e^{2T-T^2}}{ \sqrt{T}}$.

\textit{Proof}. The weight on iteration $t<T$ is $w_t=\prod_{i=t}^{T-1} e^{L_1}$ and $w_T=1$. Therefore, for all iterations $t$, $\sum_{t=1}^{T} w_{t}^{2} \leq Te^{4T^2}$. In addition, $\sum_{t=1}^{T} w_{t} \geq Te^{\frac{T(T+1)}{2}} \geq Te^{T^2}$.

Through Eq.\ref{eq6} and Lemma 2, we find that $ Q_{i}^{w, T} \leq \frac{\Delta \sqrt{|A|} \sqrt{\sum_{t=1}^{T} w_{t}^{2}}}{\sum_{t=1}^{T} w_{t}}\leq \frac{\Delta \sqrt{|A|}e^{2T-T^2} }{\sqrt{T}}$. Due to $ R_i^T\leq \sum_{I \in \mathcal{I}_i} R^T(I)$ \cite{Cesa2006Prediction}, we find that $ Q_{i}^{w, T} \leq \frac{\Delta\left|\mathcal{I}_{i}\right| \sqrt{|A|}e^{2T-T^2}}{ \sqrt{T}}$. Since $R_i^{w,T} \leq Q_i^{w,T}$, thus $R_i^{w,T} \leq \frac{\Delta\left|\mathcal{I}_{i}\right| \sqrt{|A|}e^{2T-T^2}}{ \sqrt{T}}$.

It can be found that the regret bound of our method is never higher than that of the vanilla CFR. We give a brief analysis on the regret bound. $\Delta \sqrt{|A|} e^{2 T-T^{2}}/{\sqrt{T}}$ is the regret bound of our method ECFR, and $\Delta\left|I_{i}\right| \sqrt{\left|A_{i}\right|} / \sqrt{T}$ is the regret bound of the vanilla CFR \cite{zinkevich2008regret}. $R_{ECFR}$ and $R_{CFR}$ are used to represent these two regret bounds respectively. $R_{ECFR}/ R_{CFR}$ = $\Delta \sqrt{|A|} e^{2 T-T^{2}}/{\sqrt{T}}/\left(\Delta\left|I_{i}\right| \sqrt{\left|A_{i}\right|}\right) / \sqrt{T}$ = $e^{2T}/\left( e^{T^2}  \left|I_{i}\right|  \right)$. And $e^{2T}/\left( e^{T^2}  \left|I_{i}\right|  \right)$ $\leq e^{2T}/ e^{T^2} $, since $\left|I_{i}\right|$ is the number of the information set of player $i$, which is an integer and not less than 1. For $e^{2T}/ e^{T^2} $, $e^{2T}/ e^{T^2} < 1$ when $T > 2$, $T$ is the number of iteration that is an integer and not less than 1. And in general, the number of iterations is much greater than 2. Thus, we can conclude that the regret bound of our method is never higher than that of the vanilla CFR.

\subsection{Differences between ECFR and other CFR-based Methods}
In this section we analyze the differences between our ECFR method and the other three major CFR-based methods that also aim to speed up the convergence of vanilla CFR. These are, LCFR \cite{brown2019deep}, DCFR \cite{brown2019solving} and dynamic thresholding for CFR \cite{Noam2017dynamic} which we will call dynamic CFR. 

Both of LCFR and DCFR improve upon CFR by applying a reweighting strategy. DCFR \cite{brown2019solving} is implemented by discounting the immediate regret value of each iteration. For a positive immediate regret value, the regret is multiplied by a weight of $\frac{t^{\alpha}}{t^{\alpha}+1}$. For a negative immediate regret value, the regret is multiplied by the weight of $\frac{t^{\beta}}{t^{\beta}+1}$. In addition, in DCFR, the average strategy is multiplied by $\left(\frac{t}{t+1}\right)^{\gamma}$ to obtain the final strategy. After the above three forms of discount, the regret value at each iteration  and the final strategy are updated by the reallocated weight. LCFR \cite{brown2019deep} uses the iteration $t$ to weight the immediate regret value. That is to say, the regret value is weighted by the iteration $t$ at each iteration as the number of iterations increases. Dynamic CFR \cite{Noam2017dynamic}, speeds up the convergence by pruning parts of the decision tree using dynamic thresholding. 

Although our ECFR also reweights the regret value on each iteration, it is different from DCFR and LCFR in both the overall concept and in the practical implementation. First of all, our approach comes from the intuitive idea that whichever form of weight is given to the regret values, the action with a higher regret value will be given a larger probability, where the ultimate goal is to further accelerate the convergence of the strategy by reweighting the regret value. Secondly, in terms of implementation details, our method reallocates the weight via the exponential form of the $L_1$ loss, which is different from DCFR, which is implemented by discounting based on the number of iterations. Therefore, our method is distinct from DCFR, and LCFR.

In Dynamic CFR, the exponential weight is adopted in calculating the strategy of next iteration and a hedging algorithm is used to minimize regret. However, our ECFR approach is quite different from dynamic CFR in six aspects. Firstly, our approach uses exponential weight to calculate the cumulative regret and the next iteration strategy, while dynamic CFR only uses exponential weight in calculating the strategy of next iteration. Secondly, our approach gives a small value to the negative immediate regret because we believe that the actions that give rise to negative regret in the early stages are also relevant, while dynamic CFR only deals with positive regret. Thirdly, the exponential weight is only used in dynamic CFR when using the hedging  algorithm to minimize the regret. Conversely, our approach uses a regret matching algorithm as the regret minimization method for the strategy iteration. Fourthly, dynamic CFR does not traverse all nodes in the game tree, but prunes the nodes below a threshold to accelerate the convergence. Our approach traverses all the nodes in the game tree, and accelerates convergence by redistributing the weight of regret. Fifthly, the parameter settings for exponential weighting are different in the two methods. In dynamic CFR, $\frac{\sqrt{\ln (|A(I)|)}}{3 \sqrt{\operatorname{VAR}(I)_{t}} \sqrt{t}}$ is set (VAR$(I)_t$ is the observed variance of $v(I)$ up to iteration $t$), while in our approach $L_1$ is set ($L_1 = r^i_{(I,a)} - EV_I$). Finally, from the theoretical analysis, the regret boundary of the two methods is different. The regret boundary of dynamic CFR is $R^{T}(I) \leq C \sqrt{2} \Delta(I) \sqrt{\ln (|A(I)|)} \sqrt{T}$ ($C\geq 1$ on every iteration $t$), while the regret boundary in our method is $R^{T}(I) \leq \frac{\Delta \sqrt{|A|} e^{2 T-T^{2}}}{\sqrt{T}}$.

\section{Experiment}
\subsection{Experimental Setup}
In recent years, the game of poker has been widely used to test the performance of CFR-based methods because it contains all the elements of an IIG. In this paper, we compare our ECFR with other CFR-based methods in the context of three different poker games, which are Kuhn, Leduc, and Royal. These three poker games are all simplified versions of Texas Hold'em poker and a popular benchmark in the IIG. We chose them as the test platform, because they are large enough to be highly nontrivial but small enough to be solvable. It can conveniently evaluate the performance of the algorithm.

It is worth noting that the three kinds of poker we used in the experiment are two-player games. Among these three kinds of poker, Kuhn poker is the simplest. Kuhn poker has three cards in total, only one round, each player has one hand, and there are no public cards. Leduc poker has six cards and operates over two rounds. Each player has a private hand in the first round and a public card in the second round. There are eight cards in Royal poker, and there are three rounds. In the first round, each player receives a private card, and a public card is issued in the second and third rounds. Tab.~\ref{tab:2} gives some more details of the three types of poker.
\begin{table*}[!t]
	\renewcommand{\arraystretch}{1.3}
	\caption{Details of the three kinds of poker}
	\centering
	\label{tab:2}       
	\begin{tabular}{|c||c||c||c||c||c||c|}
		\hline
		\textbf{Poker}&	\textbf{Total cards}&	\textbf{Public cards}&	\textbf{Private cards}&	\textbf{Round}&	\textbf{Ante}&	\textbf{Betsize} \\
		\hline \hline
		Kuhn&	3&	0&	1&	1	&1&	1\\ \hline
		Leduc&	6&	1&	1&	2&	1&	2; 4\\ \hline
		Royal&	8&	2&	1&	3&	1&	2; 4; 4\\ 
	\hline
	\end{tabular}
\end{table*}


\subsection{Experimental Results}
Exploitability is a standard evaluation metric used to measure the effectiveness of a strategy solved by an algorithm. The exploitability of a strategy is defined as: $e(\sigma_i)=u_i(\sigma_i^*,BR(\sigma_i^*))-u_i(\sigma_i,BR(\sigma_i))$, which determines how close $\sigma$ is to an equilibrium, where a lower exploitability indicates a better strategy. When the exploitability is zero, this means that the strategy cannot be beaten by any other strategy.

Two groups of experiments were conducted. The first group aimed to verify the effectiveness of our method on the three different games. The second group was an ablation study, which analyzed the effect of the results to the parameter settings.
\subsubsection{Comparison with State-of-the-art Methods}
\begin{figure*}[!t]
	\centering
	\subfloat[Tested on Kuhn]{\includegraphics[width=1\linewidth]{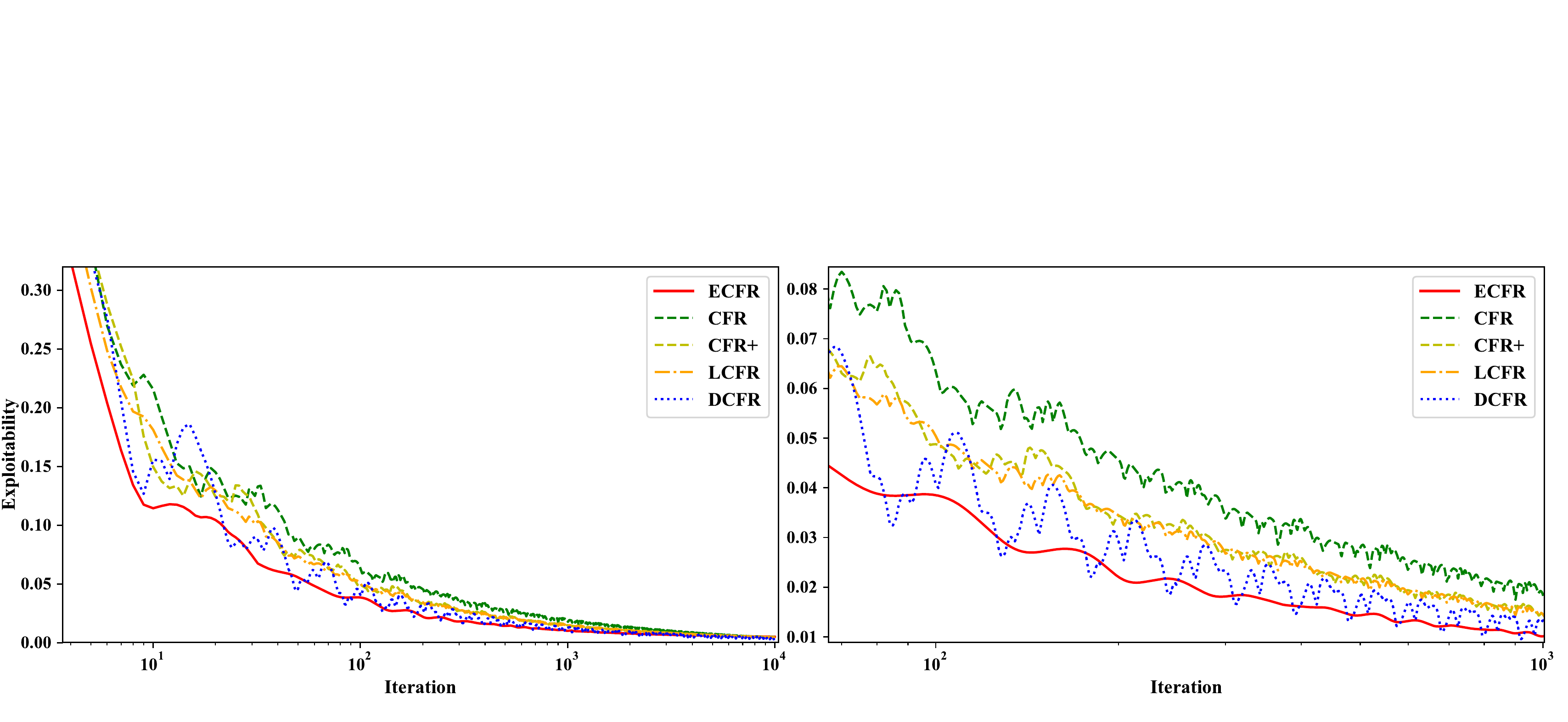}%
		\label{result1}}
	\hfil
	\subfloat[Tested on Leduc]{\includegraphics[width=1\linewidth]{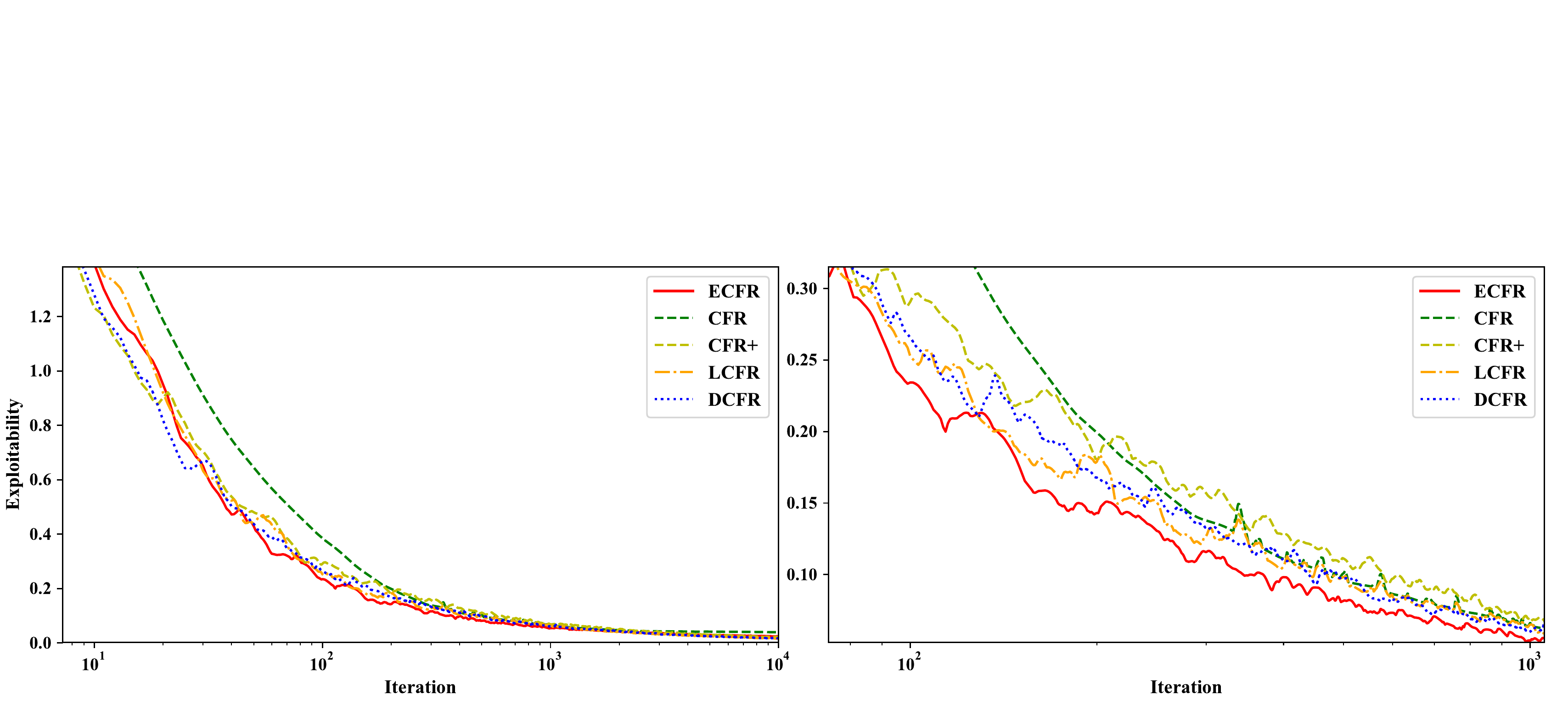}%
		\label{result2}}
	\hfil
	\subfloat[Tested on Royal]{\includegraphics[width=1\linewidth]{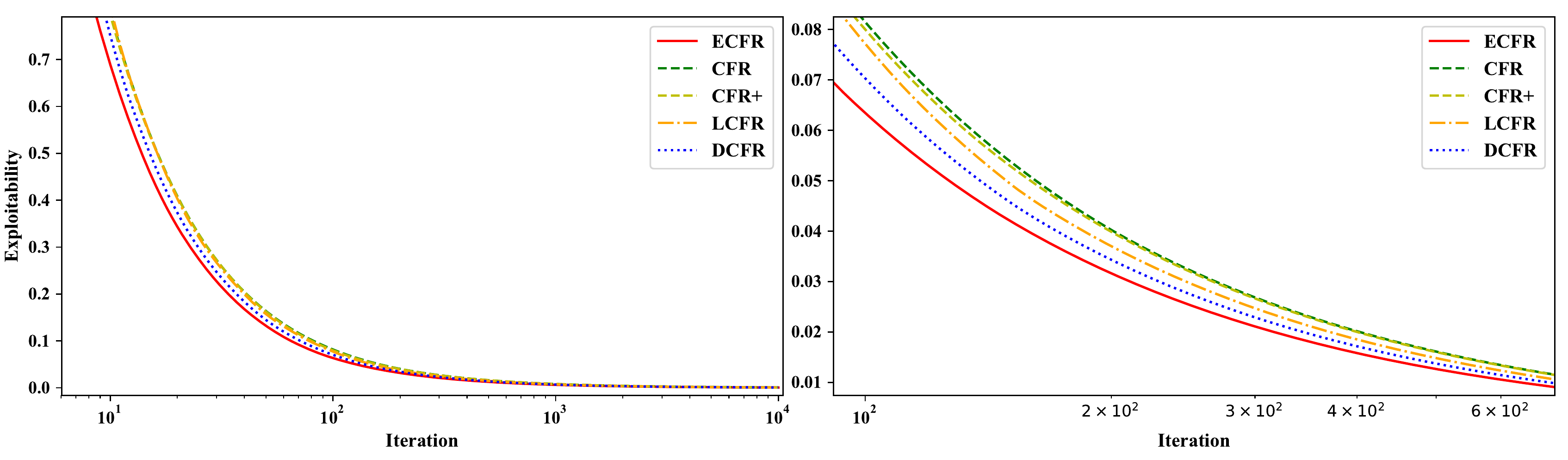}%
		\label{result3}}
	\hfil
	\caption{Comparison with state-of-the-art methods. (The Y-axis represents the exploitability, and the X-axis represents the number of iterations. The lower of the exploitability, the better the strategy is. The subfigures in the left hand column are the overall curve of the experimental results. The subfigures in the right column are a more detailed representation of parts of the subfigures in the left column, in order to better display the experimental results.)}
	\label{result}
\end{figure*}

We conducted the first group experiments with four state-of-the-art methods, which are CFR \cite{zinkevich2008regret}, CFR+ \cite{bowling2015heads}, LCFR \cite{brown2019deep}, and DCFR \cite{brown2019solving} respectively. We found that, for all the methods, when the number of iterations reached 10,000, the reduction in exploitability became very small. Thus, we limited the tests to 10,000 iterations, which we believe is sufficient to demonstrate the effectiveness of each method. The experimental results are shown in Fig.~\ref{result}. Note that the sub-figures in the left hand column show the overall progress of the experiments, while the sub-figures in the right hand column highlight some details of the sub-figures in the other column.

As shown in Fig.~\ref{result}, four methods were tested in the experiments. CFR \cite{zinkevich2008regret}, CFR+ \cite{bowling2015heads}, LCFR \cite{brown2019deep}, and DCFR \cite{brown2019solving} were used for comparison with our ECFR approach. CFR \cite{zinkevich2008regret} was the first method to solve the strategy through regret matching in IIGs. CFR+ uses regret-matching+ to update the strategy of the next iteration, which converges faster than CFR. LCFR and DCFR are both CFR-based methods which reweight regrets from iterations in various ways. Among them, we chose the parameters $\alpha=1.5$, $\beta=0$, and $\gamma=2$ in the DCFR, which were the optimal parameters given in \cite{brown2019solving}. 

First, we analyzed the convergence of our approach. In Fig.~\ref{result} (especially in the subfigures on the left), we found that our ECFR approach (indicated by the solid red line) always ended up close to zero in the three test games. In addition, from the overall trend of the curve, we also find that, with increasing iterations, ECFR shows a similar trend to the other four methods. That is, the curve presents a downward trend. Moreover, the convergence of the ECFR has been proven theoretically earlier in Sec.3.3. Therefore, the convergence of our approach has been verified from both experimental and theoretical perspectives. 

Fig.~\ref{result1} shows the experimental results for Kuhn poker. It can be seen that ECFR performs better than the other methods overall, although DCFR performs better than ECFR over some ranges of iterations, such as $t=82\sim 88,114\sim 120,182\sim 185, 241\sim 248$. However,  the convergence of LCFR appears to be unstable and fluctuates a lot. In contrast, our approach performs better than the other methods in most iterations, and the performance is relatively stable.

Fig.~\ref{result2} shows the experimental results for Leduc poker. It can be seen that in iterations $t=73\sim 76, 118\sim 122$, the exploitability of ECFR (the red curve) converges slightly slower than that of DCFR and LCFR. However, apart from these limited number of iterations, our method appears to converge faster than the other methods.

Fig.~\ref{result3} shows the experimental results for Royal poker. It can be seen clearly from the figure that the performance of our method is better than that of the other comparison methods. The red curve is always at the bottom compared with the other curves, which indicates that it is converging earlier.


To sum up, three games were used to test the effectiveness of our method. In terms of convergence, the experimental results show that our method converges reliably. In terms of the rate of convergence, the experimental results show that our method can speed up the convergence of the strategy, which shows a better performance than the other comparison methods. Therefore,  results fully verify the effectiveness of our ECFR method.
\subsubsection{Ablation Study}\label{Ablation Study}
The ablation study was conducted in the second group of experiments, which includes two aspects. First, we analyzed the sensitivity of the results to the different parameter settings. Secondly, we verified the effect of parameter $\beta$ by setting with/without $\beta$ in ECFR.

\begin{figure*}[!t]
	\centering
	\subfloat[Multiple groups of different parameter settings]{\includegraphics[width=1\linewidth]{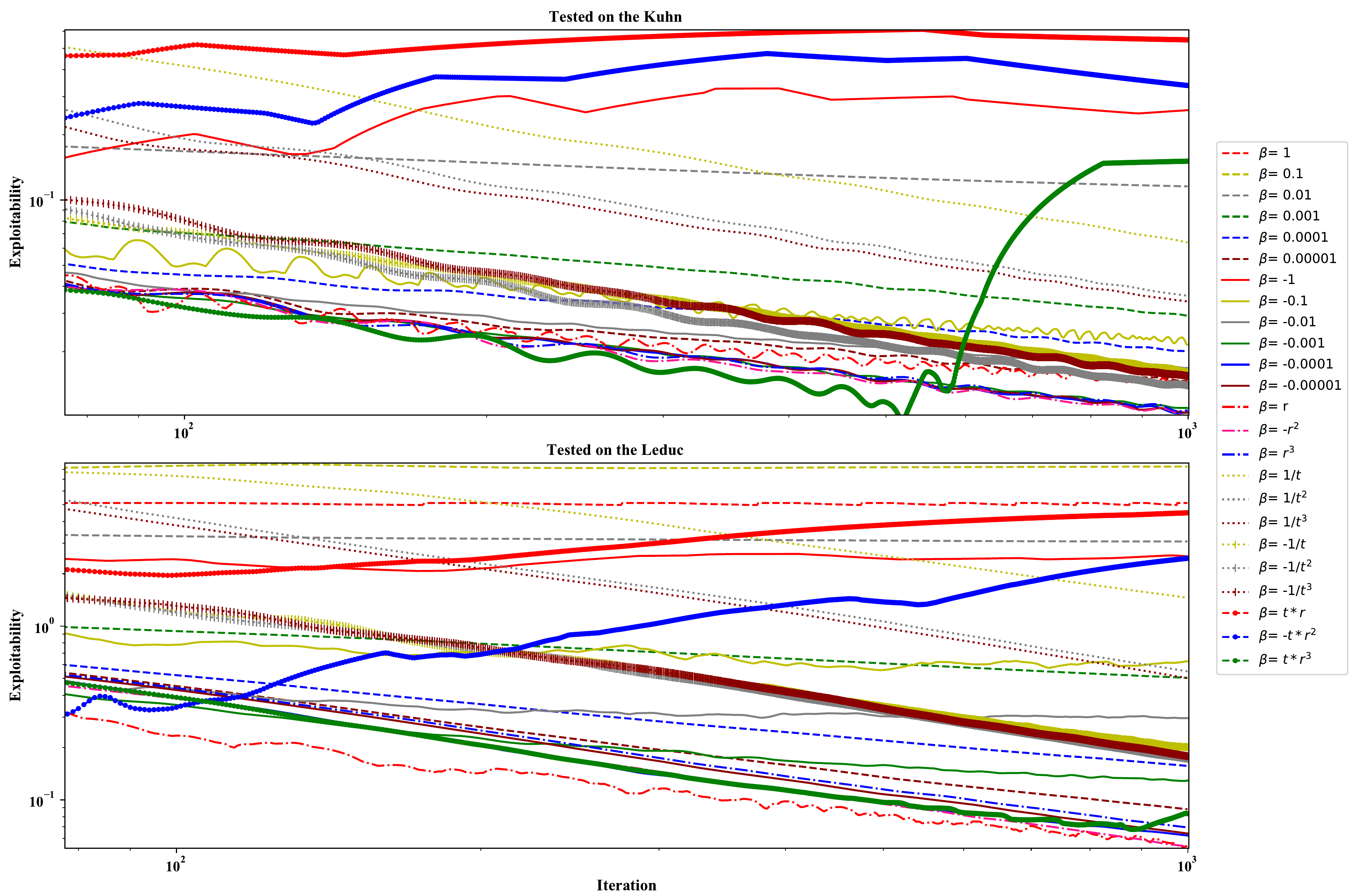}%
		\label{canshu1}}
	\hfil
	\subfloat[Fine-tuning parameter settings]{\includegraphics[width=1\linewidth]{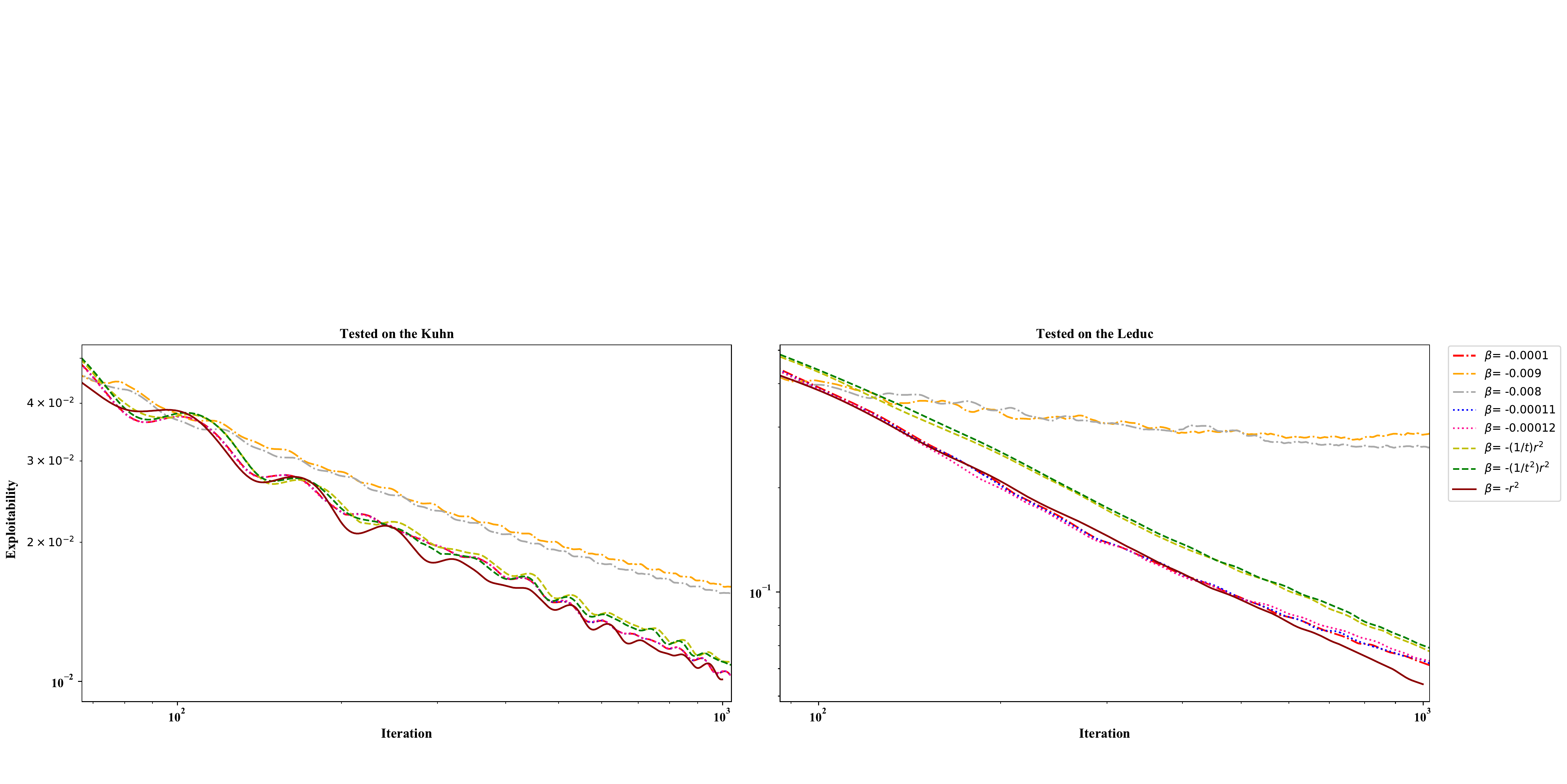}%
		\label{canshu2}}
	\caption{Ablation study on the different parameter settings. (The Y-axis represents the exploitability, and the X-axis represents the number of iterations. The smaller of the exploitability, the better. For parameter setting, we also conducted two groups of experiments. First, we selected a group of optimal settings from several groups of parameters, and then further refined the selected values.)}
	\label{canshu}
\end{figure*}

For $\beta$ in Eq.4, several different settings were tested, specifically $\beta$: $\pm1, \pm0.1, \pm0.01, \pm0.001, \pm0.0001, \pm0.00001$; $r, r^2, r^3$; $\pm \frac{1}{t}, \pm \frac{1}{t^2}, \pm \frac{1}{t^3}$; $tr, tr^2, tr^3$, where $r$ is the instantaneous regret for each action, and $t$ is the number of iterations. The results are shown in Fig.~\ref{canshu1}. We start with a rough selection of four different sets of values and then make a further, more detailed selection. For the fine-tuning selection, the value of $\beta$ was set to: $-0.008, -0.009, -0.0001, -0.00011, -0.00012, -r^2, -\frac{r^2}{t}, -\frac{r^2}{t^2}$, and the results are shown in Fig.~\ref{canshu2}. Considering that we are only choosing the optimal parameter settings here, two games (Kuhn and Leduc) were used to test the settings. The number of iterations was set to 1,000. 

As shown in Fig.~\ref{canshu1}, we find that $\beta= -0.0001$, $\beta= -r^2$, $\beta= -0.00001$, and $\beta= r^3$ have a good performance in the Kuhn poker game. In Leduc, $\beta= -0.0001$ and $\beta= -r^2$ produce a good performance, but $\beta= -0.00001$ and $\beta= r^3$ perform worse than $\beta= -0.0001$ and $\beta= -r^2$. 

On the basis of these results, $\beta= -0.0001$ and $\beta= -r^2$ were chosen for further optimization. In order to fine-tune the appropriate value for $\beta$, we set $\beta= -0.0001$, $-0.008, -0.009, -0.00011, -0.00012$. Also, $\beta= -r^2$, $-\frac{r^2}{t}$ and $-\frac{r^2}{t^2}$ were added for comparison. The results are shown in Fig.~\ref{canshu2}. In the Kuhn poker game, we found that $\beta= -r^2$ performs the best when compared against the other settings in Fig.~\ref{canshu2}. In the Leduc poker game, although $\beta=-0.0001,0.00011,0.00012$ perform better in the first 400 iterations, the performance of $\beta= -r^2$ gradually exceeds the others after that point. Therefore, $\beta= -r^2$ was selected as the final setting in the comparison experiments.

In addition, we also analyzed the effect of $\beta$ to the algorithm performance. Based on the analysis of Section Sect.\ref{ewt}, we think the $\beta$ can contribute to the convergence of the strategy. Here we conducted the experiment to verify the effect of $\beta$ by setting with/without $\beta$ in ECFR. To better reflect fairness, we set the number of iterations to 10000.

\begin{figure*}[!t]
	\centering
	\includegraphics[width=1\linewidth]{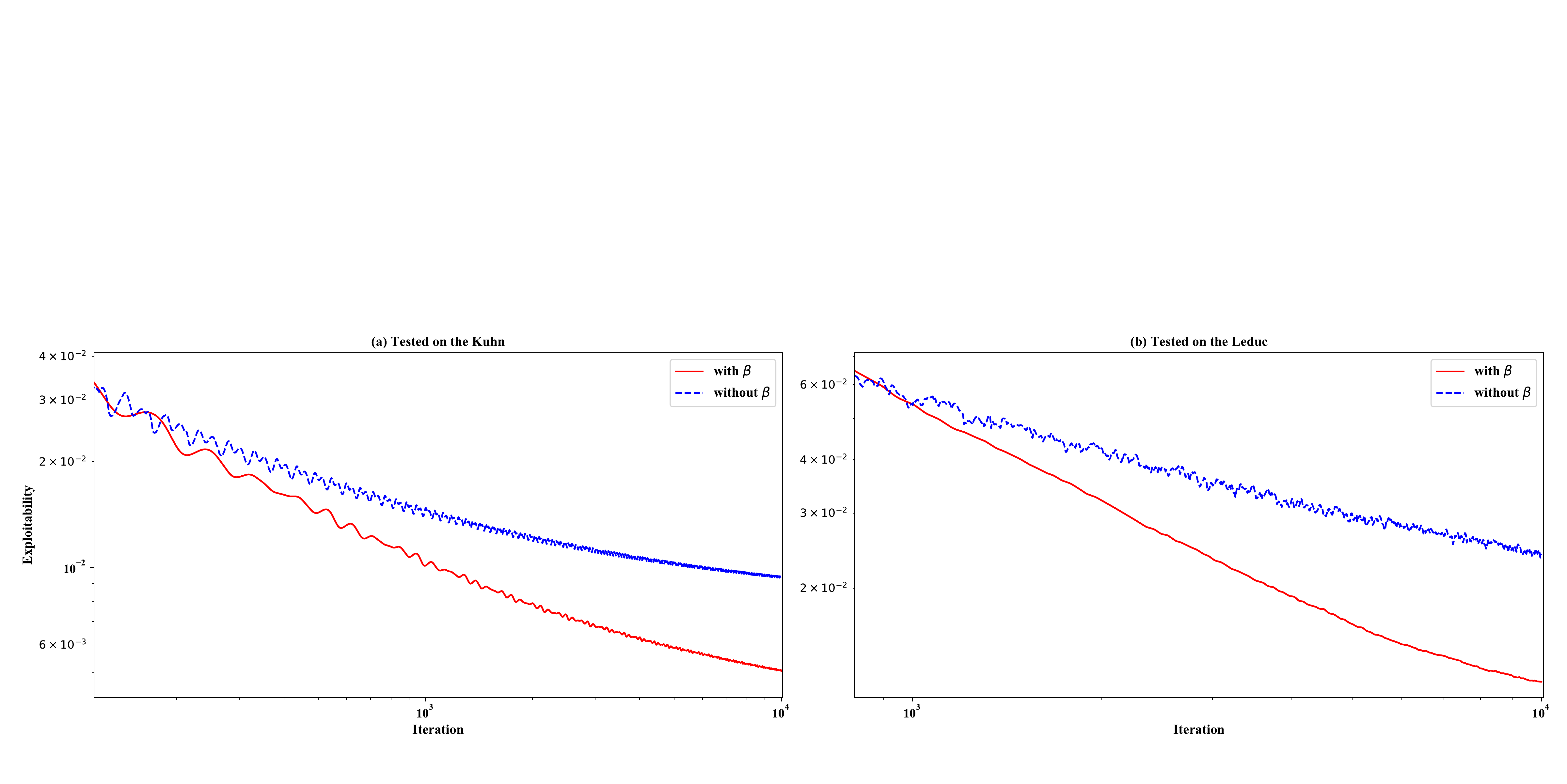}
	\caption{Ablation study on the parameter with/without $\beta$. (The Y-axis represents the exploitability, and the X-axis represents the number of iterations. The smaller of the exploitability, the better.)}
	\label{xiaorong}       
\end{figure*}

The results are shown in Fig.~\ref{xiaorong}. In the Kuhn poker, we found that the performance of 'with $\beta$' completely exceeds  'without $\beta$' after the first 100 iterations. In the Leduc poker, although 'without $\beta$' performs better in the first 750 iterations, the performance 'with $\beta$' gradually exceeds the other after that point. The experimental results fully verify that the setting of parameter $\beta$ is effective in the ECFR.

\section{Conclusion}
In this paper we proposed an exponential counterfactual regret minimization algorithm named ECFR. It can be used to build the approximate Nash equilibrium strategy of an extensiveform imperfect information game. We introduce the exponential weighting technique for regret in the process of iteration, and provide a detailed theoretical proof of convergence. Extensive experiments were then conducted on three kinds of game. Under the same number of iterations, the exploitability of the strategy obtained by our method is the lowest. The results demonstrate that our method not only has a good convergence, but also converges faster than current state-of-the-art methods.

\bibliographystyle{IEEEtran}
\bibliography{IEEEabrv,yourconference2}
\end{document}